\documentclass[twocolumn,amsmath,amssymb,footinbib,superscriptaddress]{revtex4-2}

\usepackage{graphicx,color}
\usepackage{dcolumn}
\usepackage{bm}
\usepackage{here}

\usepackage{epstopdf}
\usepackage{amsmath}
\usepackage[colorlinks=true,citecolor=blue,urlcolor=blue,linkcolor=blue]{hyperref}

\begin{document}

\title{Robust coherent phonon mode at GaP/Si(001) heterointerface}

\author{Kunie Ishioka}
\email{ishioka.kunie@nims.go.jp}
\affiliation{National Institute for Materials Science, Tsukuba, 305-0047 Japan}

\author{Gerson Mette}
\affiliation{Faculty of Physics and Materials Sciences Center, Philipps-Universit{\"a}t Marburg, 35032 Marburg, Germany}

\author{Steven Youngkin}
\affiliation{Faculty of Physics and Materials Sciences Center, Philipps-Universit{\"a}t Marburg, 35032 Marburg, Germany}

\author{Andreas Beyer}
\affiliation{Faculty of Physics and Materials Sciences Center, Philipps-Universit{\"a}t Marburg, 35032 Marburg, Germany}

\author{Wolfgang Stolz}
\affiliation{Faculty of Physics and Materials Sciences Center, Philipps-Universit{\"a}t Marburg, 35032 Marburg, Germany}

\author{Kerstin Volz}
\affiliation{Faculty of Physics and Materials Sciences Center, Philipps-Universit{\"a}t Marburg, 35032 Marburg, Germany}

\author{Christopher J. Stanton}
\affiliation{Department of Physics, University of Florida, Gainesville, FL 32611 USA}

\author{Ulrich H{\"o}fer}
\affiliation{Faculty of Physics and Materials Sciences Center, Philipps-Universit{\"a}t Marburg, 35032 Marburg, Germany}

\date{\today}

\begin{abstract}
Lattice-matched GaP layers without extended defects can be grown on Si(001) substrate via a two-step growth procedure, consisting of low-temperature nucleation followed by high-temperature overgrowth.  A transient reflectivity experiment on a thin, low-temperature nucleation layer discovered a previously unknown phonon mode at 2 THz upon below-bandgap optical excitation (\textit{Adv. Mater. Interfaces} \textbf{2025}, 2400573).  Here we examine the influence of the two-step growth process on the ultrafast carrier and phonon dynamics of the GaP/Si interface.  We find that the discrete electronic state, which governed the interfacial carrier dynamics of the thin nucleation layer, becomes suppressed when a thicker layer is formed by high-temperature overgrowth.  The coherent 2-THz oscillation is observed also in the high-temperature overgrown structures, at the constant frequency regardless of the GaP layer thickness.  Its resonance behavior closely follows that of the carrier dynamics at the respective growth stage.  This supports its assignment to a phonon mode generated at the heterointerface and strongly coupled to the interfacial carriers.  The phonon amplitude exhibits a non-monotonic dependence on the GaP layer thickness, and its optical polarization-dependence is qualitatively altered by the high-temperature overgrowth, neither of which is accounted for by the carrier-phonon coupling alone.  Our results demonstrate that the 2-THz interfacial phonon mode is robust against high-temperature overgrowth, while its amplitude is determined by both coupling to interfacial electronic transitions and atomic-scale structural reorganization at the interface.
\end{abstract}

\maketitle

\section{INTRODUCTION}

The electron–phonon interaction is a key factor governing charge carrier transport at buried interfaces in semiconductor devices.   At such interfaces, the breakdown of translational symmetry can give rise to spatially localized electronic states and phonon modes  \cite{Masri1988}, while also modifying the selection rules for optical excitation. Recent advances in scanning transmission electron microscopy (STEM) combined with electron energy loss spectroscopy (EELS) have enabled the mapping of phonon dispersion with sub-nanometer spatial resolution \cite{Qi2021, Cheng2021, Kikkawa2021,Li2022}.  
 Linear and nonlinear optical spectroscopies have been widely used as conventional techniques to examine the phonons at deeply buried solid-solid interfaces.  
Linear and nonlinear optical spectroscopies provide more conventional approaches for probing phonons at deeply buried solid–solid interfaces.  Among these, second harmonic generation (SHG) is particularly well suited for interface studies because it is forbidden in the bulk of centrosymmetric crystals  \cite{Matsumoto2006}. However, detecting phonons localized at a \emph{single} buried interface remains challenging due to their inherently weak signal intensity.  Optical reflection, in contrast, occurs at surfaces and interfaces where the refractive index changes abruptly, making it a suitable technique for investigating interface phonons.  Its sensitivity can be further enhanced when the phonon response is resonantly amplified through coupling with interfacial electronic transitions \cite{Mette2025}

\begin{figure*}
\includegraphics[width=0.43\textwidth]{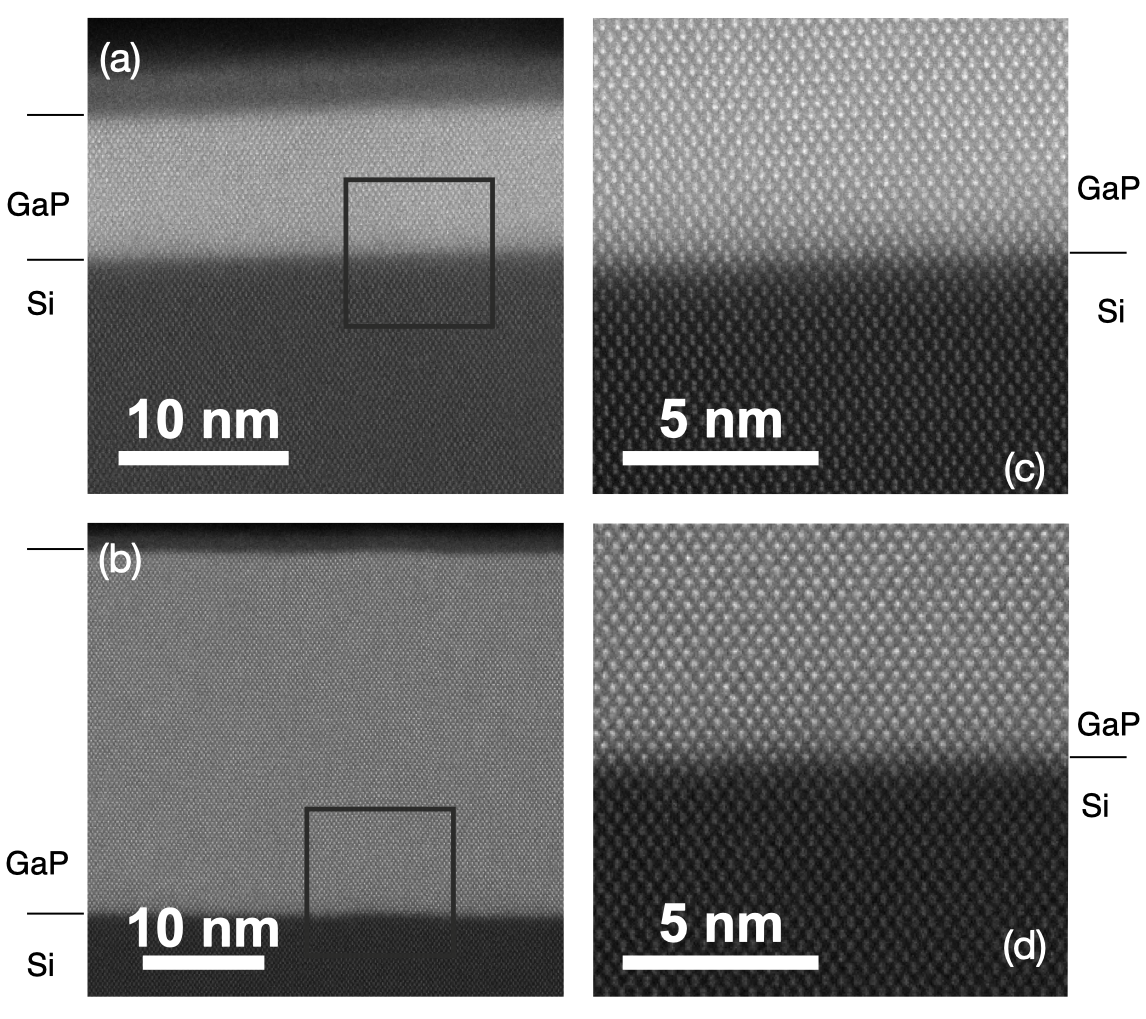}
\includegraphics[width=0.52\textwidth]{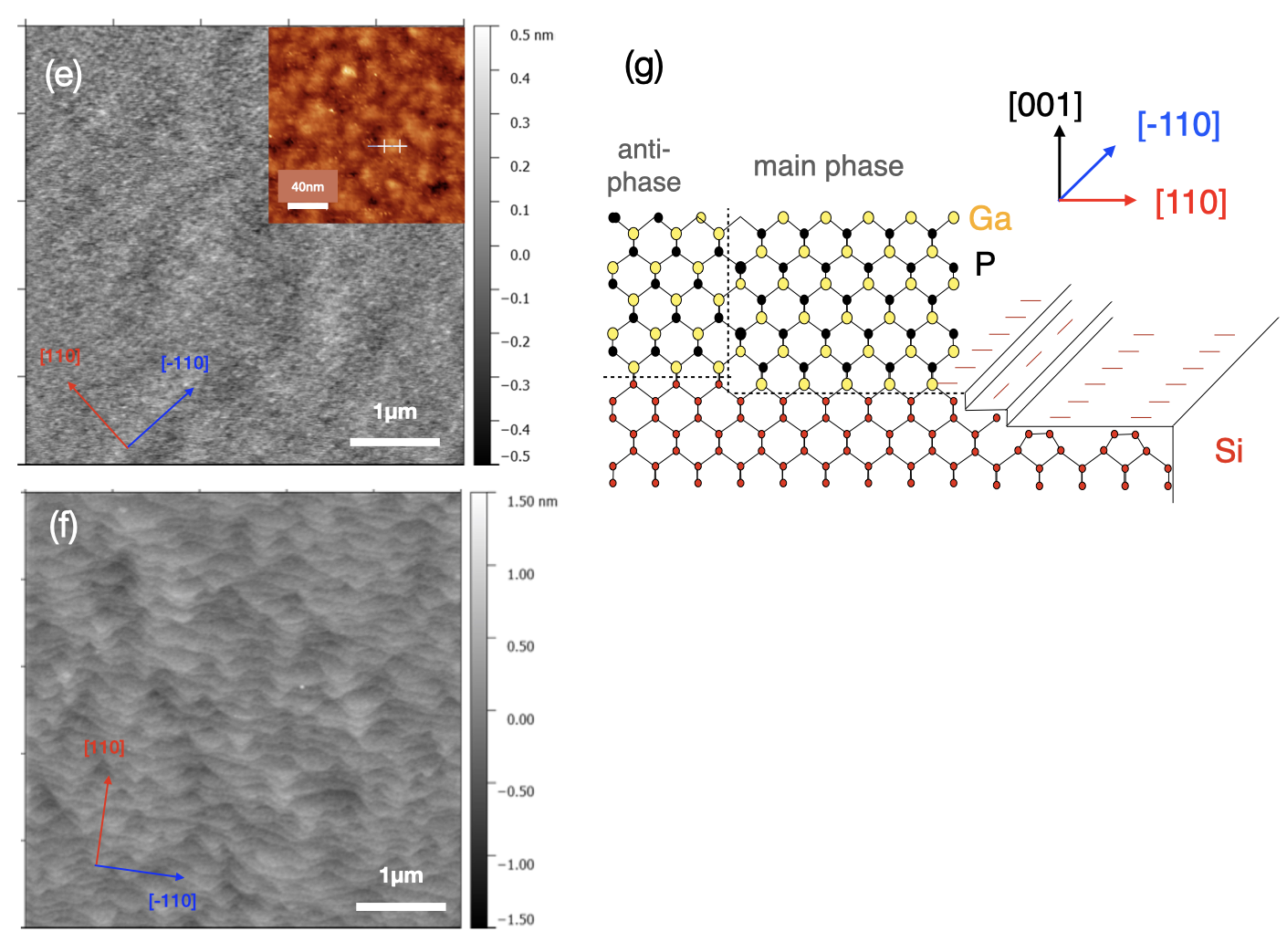}
\caption{\label{AFM} (a-d) HAADF-STEM images of $d=10$~nm nucleation layer F (a,c) and $d=38$~nm overgrown layer D (b,d).  Black squares in (a,b) represent the areas that are magnified in (c,d). 
(e,f) Atomic force microscopy (AFM) images of the GaP/Si(001) samples: $d=8$~nm nucleation layer A (e) and $d=48$~nm overgrown layer E (f). (g) Ball and stick model of an abrupt, Ga-terminated GaP/Si(001) interface, with the main phase and antiphase of GaP.  Red, yellow and black circles represent  silicon, gallium and phosphorus atoms, respectively.}
\end{figure*}

Lattice matched GaP layers without extended defects can be fabricated up to the thickness of $d=60$~nm on an exact Si(001) substrate via metal organic vapor phase epitaxy (MOVPE).  The standard growth procedure consists of two steps, i.e., nucleation at a low temperature followed by overgrowth at a higher temperature \cite{Lin2013, Supplie2014, Volz2011, Beyer2011, Beyer2012}.  
Density functional theory (DFT) simulations \cite{Supplie2014, Romanyuk2016, Beyer2016} predicted that atomically abrupt interfaces to be energetically less favorable than intermixed interfaces.  The metastability of the abrupt interface is experimentally confirmed by scanning transmission electron microscope (STEM) studies that revealed the formation of an intermixing layer including pyramidal nano-facets after the overgrowth \cite{Beyer2016, Beyer2019}. 
One of the DFT studies \cite{Romanyuk2016} also predicted that both atomically abrupt and intermixed GaP/Si interfaces can accommodate localized electronic states.  
Previous time-resolved SHG \cite{Mette2020} and transient reflectivity (TR) \cite{Mette2025} experiments using near infrared (NIR) pulses confirmed the existence of such an interface electronic state by observing a resonance enhancement of ultrafast carrier responseat photon energy of 1.4~eV, which was short of the bandgap of GaP.  Furthermore, the TR study revealed a previously unknown periodic modulation at 2 THz, whose resonance behavior traced that of the interface carrier excitation and thereby indicated its origin as a phonon mode localized at the heterointerface \cite{Mette2025}.  
These results were in drastic contrast to the TR experiments using near ultraviolet (NUV) pulses \cite{Ishioka2016, Ishioka2017, Ishioka2019}, where the carrier and phonon dynamics of the GaP/Si interface were roughly similar to those of the respective bulk semiconductors upon above-bandgap excitation.  
Both the SHG and TR studies using NIR pulses focused on the interface of a thin ($d\leq10$~nm) low-temperature grown nucleation layer, however.  The electronic states and phonons at the interface of thicker high-temperature overgrown layers are yet to be examined.

In this study, we investigate the effect of two-step growth process on ultrafast carrier and phonon dynamics at the GaP/Si(001) interface. TR measurements with below-bandgap excitation of GaP are performed as functions of the GaP layer thickness, pump/probe polarization, and pump photon energy. The interface electronic state of the nucleation layer is found to be quenched by the high-temperature overgrowth. On the other hand, the 2-THz phonon mode appears in the thick GaP overgrown layers as well as in the thin nucleation layers.  Whereas the resonance behavior of the phonon mode indicates its strong coupling with the interface electronic transition, other results cannot be explained solely by the electron-phonon interaction and suggests the influence of atomic reorganization at the heterointerface induced by the high-temperature overgrowth.  

\section{EXPERIMENTAL methods and data analyses}

\begin{table}
\caption{\label{Sample} The total thickness ($d$) of the GaP layer, together with the thicknesses achieved in the first low-temperature nucleation step and in the second high-temperature overgrowth step, of the GaP/Si(001) samples studied. }
\begin{ruledtabular}
\begin{tabular}{llccc}
Sample&total $d$&low temp.&high temp.\\
&(nm)&(nm)&(nm)\\
\hline
A&8&8&0\\
B&18&8&10\\
C&28&8&20\\
D&38&8&30\\
E&48&8&40\\
F&10&10&0
\end{tabular}
\end{ruledtabular}
\end{table}

The samples studied are nominally undoped GaP layers grown by metal organic vapor phase epitaxy (MOVPE) on an  $n$-type ($\rho=0.007-0.02~\Omega$cm) Si(001) substrate with a small ($0.17-0.33^\circ$) intentional miscut in the [110] direction.  
The growth of the GaP layers follows the standard two-step procedure \cite{Volz2011, Beyer2012}.   In the first step a GaP nucleation layer is grown to the thickness of 8 nm on a Si homoepitaxial buffer in a flow-rate modulated epitaxy at 450$^\circ$C in order to achieve a charge neutral interface and two-dimensional growth.  In the second step, a GaP layer is overgrown on top of the nucleation layer in continuous epitaxy at 675$^\circ$C for different temporal durations to achieve different thicknesses.  We also examine a 10-nm thick nucleation layer for comparison, whose ultrafast electron-phonon coupling was reported previously \cite{Mette2025}.  

Aberration corrected high angle annular dark field scanning transmission electron microscopy (HAADF-STEM) measurements are carried out in a JEOL JEM 2200FS operating at 200kV on the GaP/Si interfaces.  The zoom-out images [Fig.~\ref{AFM}(a,b)] confirm the total thickness $d$ of each sample given in Table~\ref{Sample}, whereas the zoom-in images [Fig.~\ref{AFM}(c,d)] emphasize that both the nucleation layer and the overgrown layer are single crystals without extended defects.  Atomic force microscope (AFM) images show that the top surface of the nucleation layer is uneven with nano-mounds of $\sim$20 nm lateral size and $\sim$1 nm height [Fig.~\ref{AFM}(e)], whereas that of the overgrown GaP layer shows domains that directly reflect the steps and terraces of the underlying Si substrate [Fig.~\ref{AFM}(f)].   

One- and two-color pump-probe reflectivity measurements are performed in a near back-reflection geometry in ambient condition. Pump-photon energies used in the present study are smaller than the indirect band gap of GaP (2.26 eV) but comparable with or larger than that of Si (1.12 eV). 
The one-color measurements, whose details are described elsewhere \cite{Ishioka2022}, are performed in both slow-scan and fast-scan schemes.  In the slow-scan experiments an output of a regenerative amplifier with 120 fs duration, 810 nm wavelength, and 100 kHz repetition rate is used as the light source.  Pump-induced change in the reflectivity $\Delta R$ is measured as the time delay $t$ between the pump and probe pulses is varied step by step with a linear motor stage.  In the fast scan experiments an output of a Ti:sapphire oscillator with 12-fs duration, 815-nm center wavelength and 80-MHz repetition rate is used as the light source.  $\Delta R$ is averaged in a digital oscilloscope while $t$ is scanned continuously with a fast scan delay.
The setup for the two-color measurements is similar to that employed in Ref.~\onlinecite{Mette2025}.  An output of a non-collinear optical parametric amplifier with 720 to 920-nm center wavelength (1.35 to 1.72-eV photon energy), 30-fs duration and 100-kHz repetition rate is mainly used as pump, whereas the output of another optical parametric amplifier with 800-nm center wavelength (1.55 eV photon energy), 40-fs duration and 100-kHz repetition rate is mainly used as probe.  
$\Delta R$ is measured while $t$ is scanned incrementally (slow scan).

\begin{figure}
\includegraphics[width=0.475\textwidth]{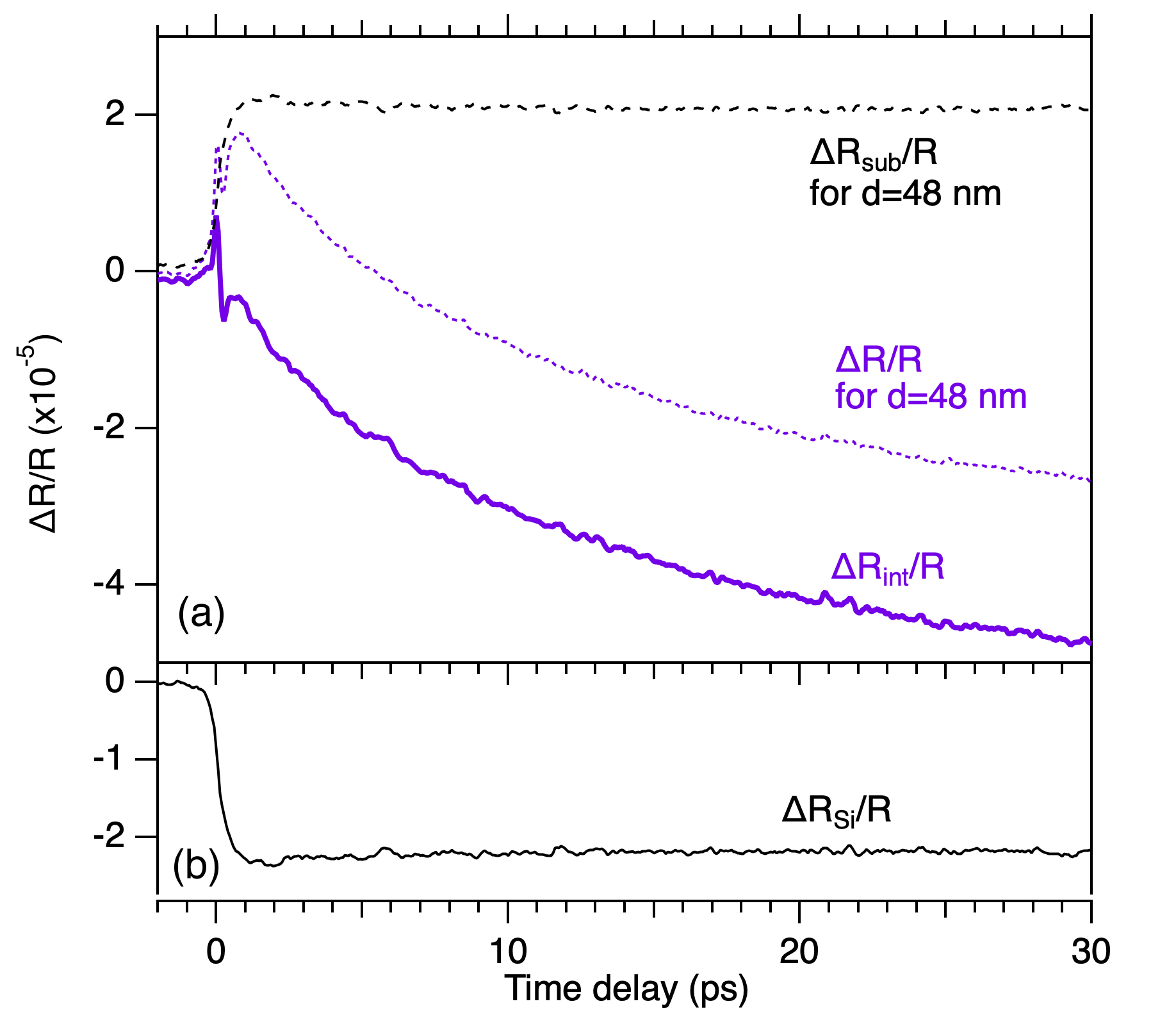}
\caption{\label{subtract} (a) As-measured TR signal (dotted trace) of a GaP layer on Si (sample F), the substrate-contribution (dashed trace) obtained from  Eq.~(\ref{sub}), and the interface-contribution (solid trace) obtained from Eq.~(\ref{E1}).  Pump and probe wavelengths are 810 nm.  (b) TR signal of blank Si substrate measured at the same condition as the GaP/Si sample.}
\end{figure}

\begin{figure}
\includegraphics[width=0.475\textwidth]{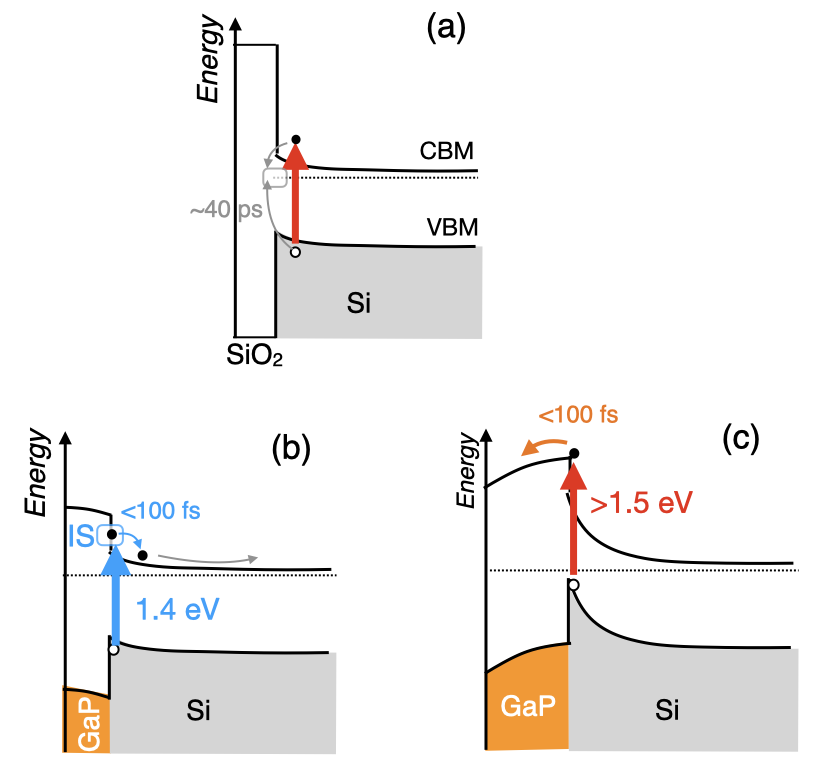}
\caption{\label{banddiagram} Schematic band energy diagrams before photoexcitation at the interface of SiO$_2$/Si (a), the GaP nucleation layer /Si (b), and GaP overgrown layer/Si (c).  Possible electronic transitions induced by NIR light and relaxation pathways are also indicated with arrows.  IS, CBM and VBM denote an unoccupied interface state, the conduction band minimum, and valence band maximum, respectively.  The band bending of the overgrown layer in (c) is based on Ref.~\onlinecite{Ishioka2016}, whereas that of the nucleation layer in (b) is unknown and drawn hypothetically, both of which are consistent with the present and previous \cite{Ishioka2016, Ishioka2022} experimental results.}
\end{figure}

As-measured transient reflectivity signals $\Delta R/R$ of the GaP/Si samples, whose example is shown with a dotted trace in Fig.~\ref{subtract}(a), contain the contribution from the Si substrate in the vicinity of the interface, $\Delta R_\text{sub}/R$, in addition to that from the GaP/Si heterointerface and the GaP overlayer, $\Delta R_\text{int}/R$.  We can extract the latter contribution by subtracting the former from the as-measured signal \cite{Ishioka2022}:
\begin{equation}\label{E1}
\dfrac{\Delta R_\text{int}(t,d)}{R}=\dfrac{\Delta R(t,d)}{R}-\dfrac{\Delta R_\text{sub} (t,d)}{R},
\end{equation}
whose result is plotted with a solid trace in Fig.~\ref{subtract}(a).
Here the substrate contribution is derived from the transient reflectivity of blank Si substrate $\Delta R_\text{Si}(t)$ measured in the same configuration [Fig.~\ref{subtract}(b)]:
\begin{equation}\label{sub}
\dfrac{\Delta R_\text{sub}(t,d)}{R}=\dfrac{T(d)}{T(0)}\dfrac{\frac{1}{R(d)}\frac{\partial R}{\partial n_\text{Si}}(d)}{\frac{1}{R(0)}\frac{\partial R}{\partial n_\text{Si}}(0)}\dfrac{\Delta R_\text{Si}(t)}{R}.
\end{equation}
$T(d)$ is the transmittance of the GaP/Si interface at the pump wavelength, $R(d)$, the reflectance at the probe wavelength, and $n_\text{Si}$, the refractive index of Si at the probe wavelength. 
The $d$-dependence of the coefficient in Eq.~(\ref{sub}) reproduces that of the Si LO phonon amplitude obtained from GaP/Si interfaces, as shown in Fig.~\ref{A_vs_d} in Appendix~\ref{AB}.
We note that in this subtraction procedure we ignore the effect of natural oxide on the surface of the blank Si substrate.  This can be justified because SiO$_2$/Si interface contributes only to a slow ($\sim40$~ps) partial recovery of the TR signal through the recombination of Si photocarriers \cite{Sabbah2000}, as schematically shown in Fig.~\ref{banddiagram}(a), whereas we focus on the carrier and phonon dynamics on faster time scale ($<10$~ps) in the present study.

\section{RESULTS}

\subsection{Growth process-dependence}\label{3A}

\subsubsection{carrier dynamics}\label{3A1}

\begin{figure}
\includegraphics[width=0.475\textwidth]{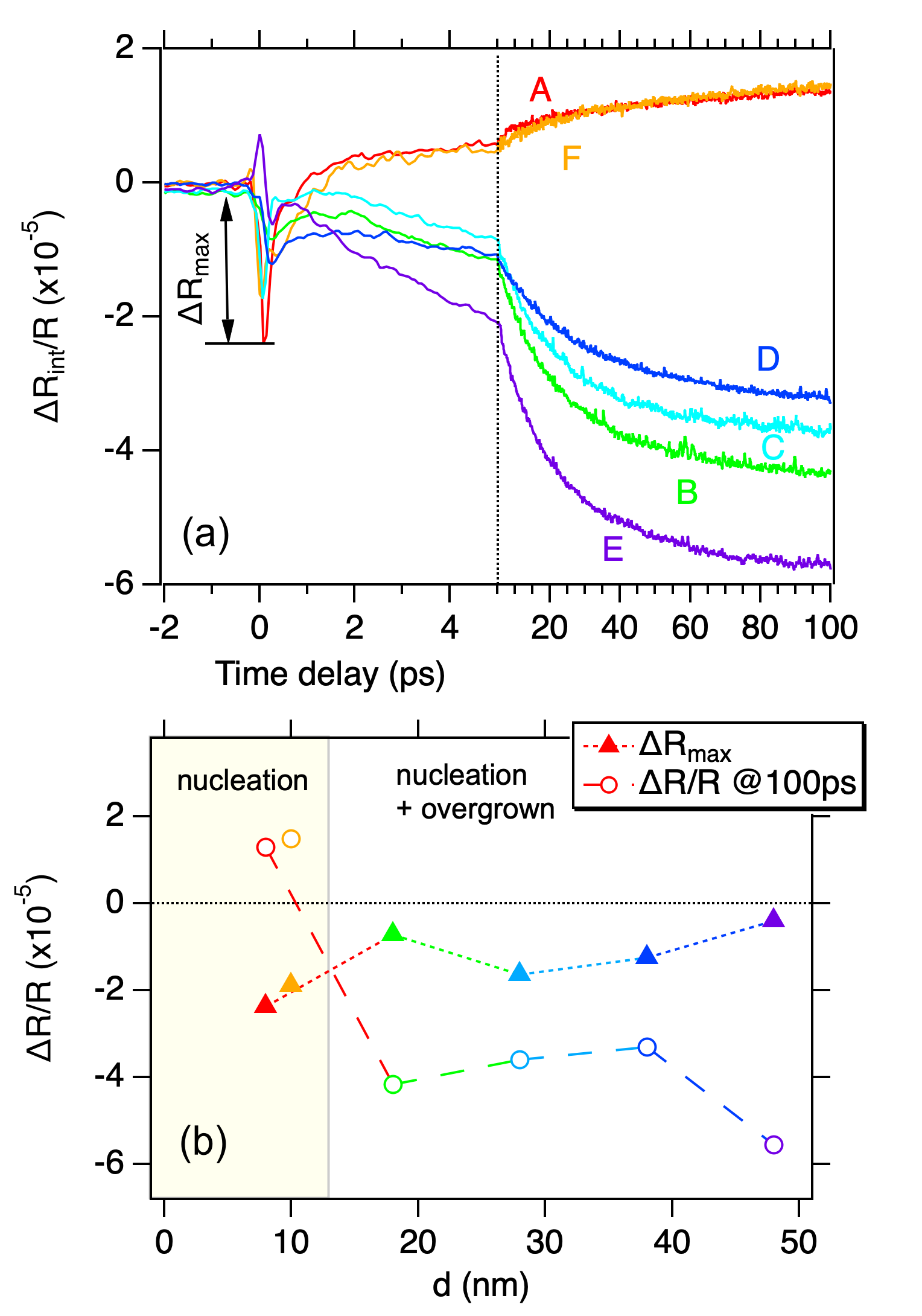}
\caption{\label{CarrierRegA} (a) Interface contribution to the slow-scan TR signals of GaP/Si samples pumped and probed at 810 nm.  Pump and probe lights are polarized along the [110] direction of the Si substrate.  Incident pump fluence is $\sim$0.25~mJ/cm$^2$. Black arrow denotes the initial drop height $\Delta R_\text{max}$.   (b) $\Delta R_\text{max}$ (filled symbols) and the TR signal at $t=100$~ps (open symbols) as a function of GaP layer thickness $d$. }
\end{figure}

We first perform one-color pump-probe measurements in the slow scan scheme using 120-fs, 810~nm laser pulses to examine the carrier dynamics of the GaP/Si interfaces.  The photon energy employed here, 1.5~eV, corresponds to the high-energy tail of the 1.4-eV resonance peak reported in the SHG and TR studies \cite{Mette2020, Mette2025}.  Figure~\ref{CarrierRegA}(a) compares the interface contribution $\Delta R_\text{int}$ to the TR signal obtained from the samples A to F. 
The pump and probe lights are both polarized along the [110] or miscut direction of the Si substrate, as schematically illustrated in Fig.~\ref{AFM}(c).  The TR signal exhibits an abrupt drop upon photoexcitation at $t=0$ for all the samples examined, indicating ultrafast creation of carriers at the heterointerface despite of the below-bandgap excitation of GaP.  The height of the initial drop $\Delta R_\text{max}$, which may be regarded as a semi-quantitative measure for photoexcitated carrier density at the heterointerface, is plotted with filled symbols in Fig.~\ref{CarrierRegA}(b).  The absolute value of $\Delta R_\text{max}$ decreases with increasing total GaP layer thickness $d$, suggesting a higher carrier density at the interface of the nucleation layer (samples A and F, $d=8$ and 10 nm) than the overgrown layer (samples B to E, $d=18$ to 48 nm).

After the initial drop, the TR signal from the nucleation layers recovers within 1 ps and continues to increase monotonically.  This temporal behavior is consistent with the previous SHG study \cite{Mette2020}, where the subpicosecond recovery and picosecond increase in the SH intensity were attributed to the relaxation of carriers photoexcited at the interface and to their diffusion into the Si bulk, respectively, as illustrated schematically in Fig.~\ref{banddiagram}(b).  
The TR signal from the overgrown layers (samples B to E, $d=18$ to 48 nm) also recovers within 1 ps after the initial drop but then turns to a decrease on tens of picosecond time scale.  The TR signal at a long time delay (e.g. $t=100$~ps) is positive for the nucleation layers and negative for the overgrown layers.  The contrast indicates that the interfacial carrier dynamics is dominated by different excitation/relaxation pathways.
A candidate for the alternative excitation pathway for the overgrown layer could be ultrafast ($<100$~fs) charge injection, either from the Si substrate, as illustrated in Fig.~\ref{banddiagram}(c), or from the interface electronic state, as  was suggested by the non-Raman generation of coherent LO phonons upon below-bandgap excitation \cite{Ishioka2022}.  
In Sect. \ref{3C} we will perform two-color TR experiments using a tunable IR pump to determine the dominant source for the charge injection.

\begin{figure*}
\includegraphics[width=0.475\textwidth]{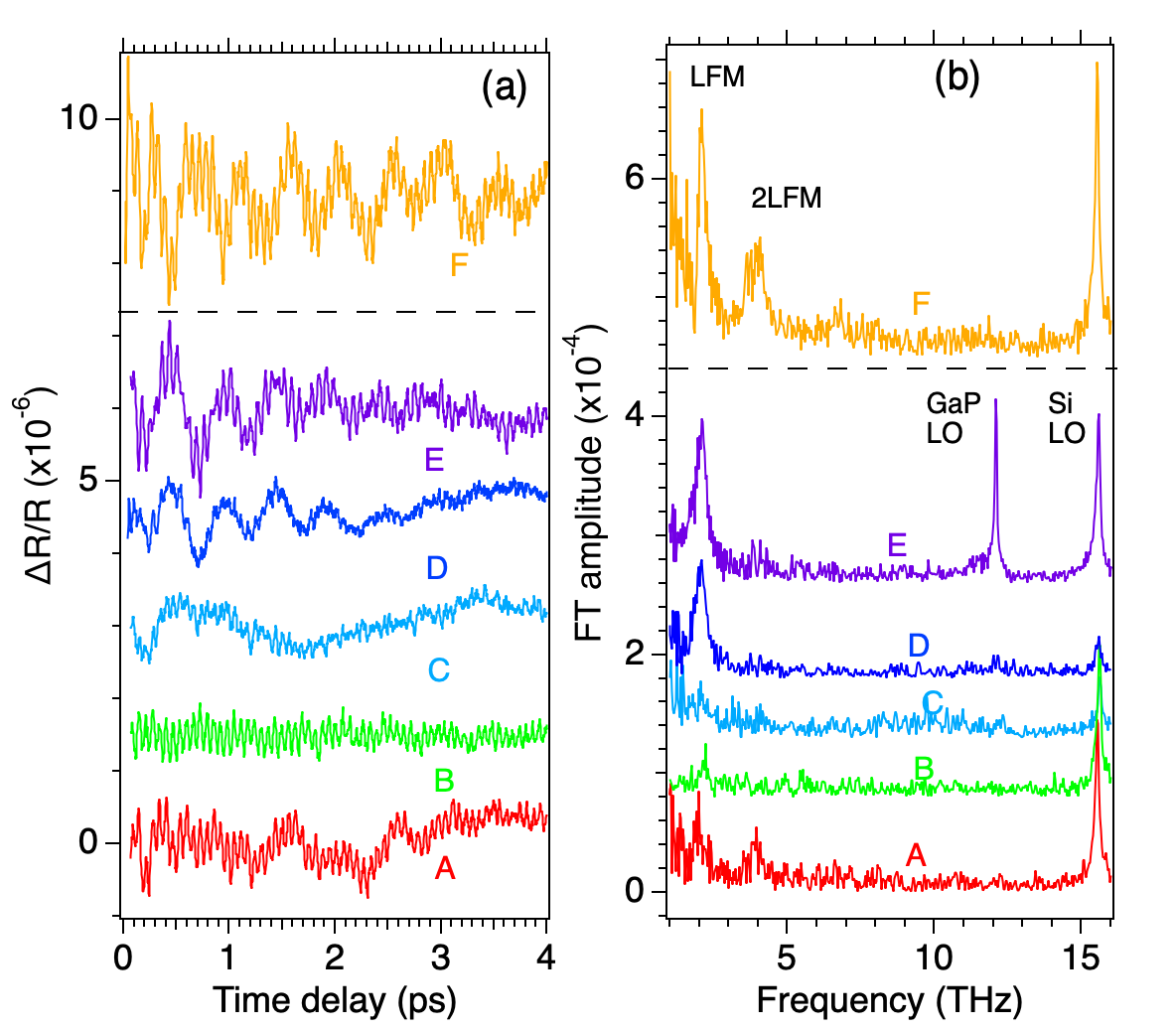}
\includegraphics[width=0.475\textwidth]{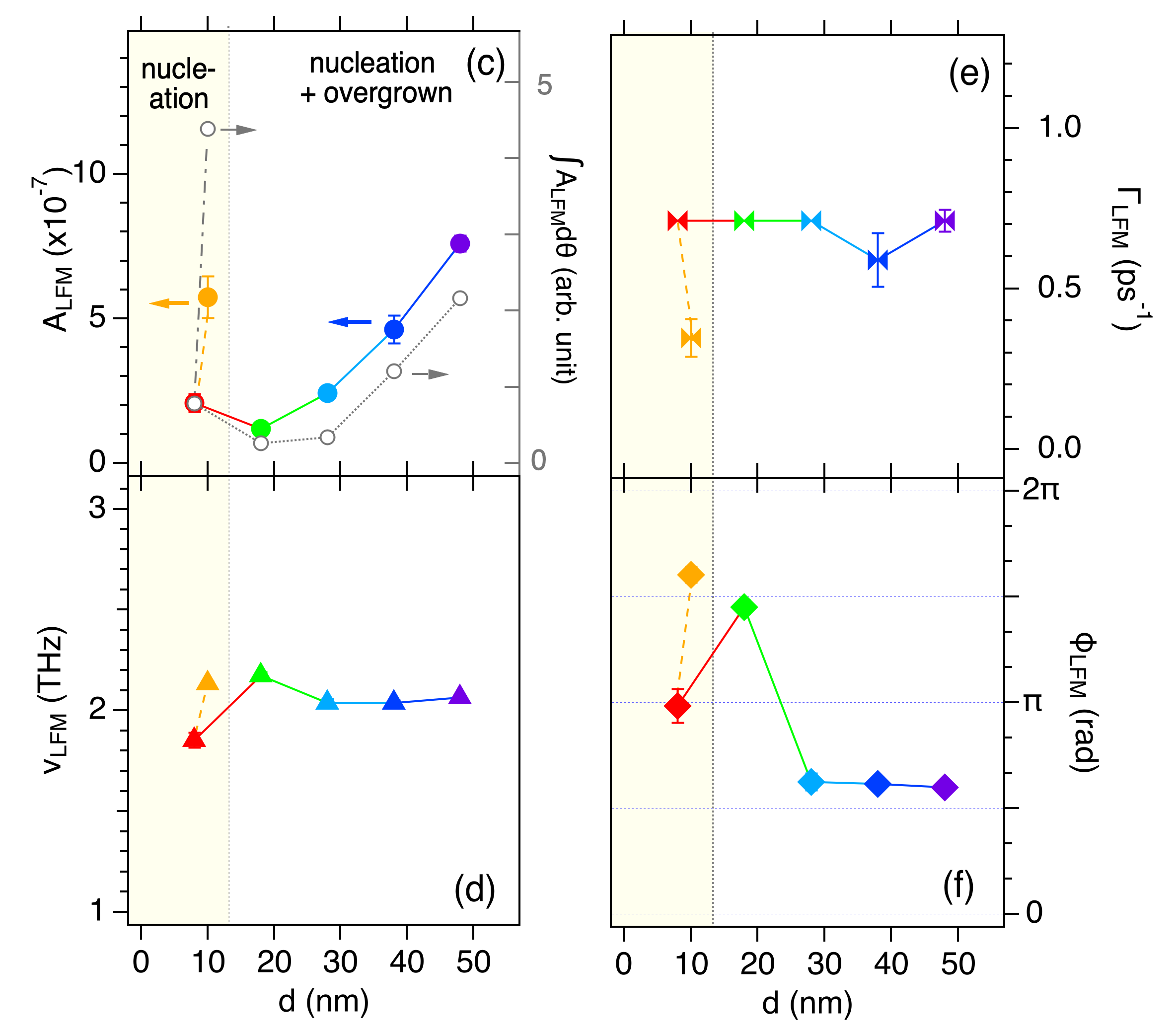}
\caption{\label{Parameters} (a)  Oscillatory part of fast-scan TR signals pumped and probed at 815 nm.  Pump and probe polarizations are parallel to the [110] direction of the Si substrate.  Incident pump fluence is $\sim0.1$~mJ/cm$^2$.  (b) Fast Fourier transform (FFT) spectra of the oscillatory TR signals in (a).  Traces are offset for clarity in (a) and (b).  (c-f) LFM amplitude (filled symbols in (c)), dephasing rate (d), frequency (e), and initial phase (f) obtained by fitting the oscillatory TR signals in (a) to Eq.~(\ref{dh}).  Open symbols in (c) represents the LFM amplitude integrated over pump angles $\theta$ from $-90^\circ$ to $90^\circ$ with probe polarized at 0$^\circ$, as described in Appendix~\ref{AC}.}
\end{figure*}

We note that the band diagram for the overgrown layer in Fig.~\ref{banddiagram}(c) is drawn based on  our previous TR study using NUV pump \cite{Ishioka2016}.  There the magnitude of the band bending was estimated from the maximum amplitude of coherent LO phonons when the built-in field was fully screened upon intense above-bandgap excitation.  Although calculation of the built-in field required a few simplifying assumptions, the obtained steep band bending can explain the efficient non-Raman generation of coherent LO phonons in the overgrown layers \cite{Ishioka2022, Ishioka2016}.  
For the nucleation layers, which was not analyzed in our previous NUV-TR study \cite{Ishioka2016}, we hypothetically draw nearly flat bands in Fig.~\ref{banddiagram}(b) so that it is consistent with the negligibly weak coherent LO phonons upon below-gap excitation \cite{Ishioka2022}.
We also note that characterization of the intrinsic electrical properties of our GaP layer by magnetotransport measurements was not straightforward, since the effective sheet resistivity was dominated by that of the n-type Si substrate rather than of the GaP layer at temperatures above 30~K \cite{Ostheim2019}.  
Secondary ion mass spectroscopy (SIMS) measurements revealed the Si impurity concentration in the GaP layer to be less than $0.1$\%  \cite{Ostheim2019}.  These results do not exclude the possibility of unintentional n-type doping comparable to that ($\lesssim10^{18}$~cm$^{-3}$) reported for a thicker ($\lesssim1\mu$m) GaP layers grown in two-step process \cite{Dixit2008}.  Yet our calculations demonstrated that the charge density affected only the curvature of the band bending, whereas the magnitude of the band bending was determined by the Fermi level pinning at the surface and at the interface, in our thin ($<50$~nm) GaP layers \cite{Ishioka2016}.
Drawing a more precise picture of the built-in field at the heterointerface would require an alternative technique such as four dimensional STEM (4DSTEM) \cite{Beyer2021, Chejaria2023, Lorenzen2024}, which is currently under planning.

\subsubsection{phonon dynamics}

Next we examine the phonon dynamics by performing one-color measurements in the fast scan scheme using 12-fs, 815-nm laser pulses, with pump and probe polarizations being parallel to the [110] direction.  Figure~\ref{Parameters}(a) compares the oscillatory part of $\Delta R$ of the GaP/Si samples after subtracting a non-oscillatory baseline.  The reflectivity oscillations feature the LO phonons of GaP and Si at 12 and 15.6 THz, whose details were reported elsewhere~\cite{Ishioka2022}.  The frequency and dephasing rate of the GaP LO phonon agree with those of bulk GaP wafer \cite{Ishioka2022}.  This, together with the absence of the forbidden transverse optical (TO) phonon at 10~THz and disorder-activated longitudinal/transverse acoustic (DALA/DATA) modes at $\sim$7 and 3~THz  \cite{Malesh1989, Azhniuk2001}, confirms the good crystallinity of the GaP layers. 

In addition to the high-frequency optical phonons, the TR signals from all the samples feature coherent low frequency mode (LFM), which was reported previously for the $d=10$~nm nucleation layer \cite{Mette2025}. The LFM is clearly seen as a peak at 2 THz in the fast Fourier transform (FFT) spectra in Fig.~\ref{Parameters}(b), together with an apparent overtone (2LFM) at 4 THz. 
The LFM is not detected in the TR signal from bulk GaP wafer or from blank Si substrate, as reported in Refs.~\onlinecite{Ishioka2022, Mette2025} and shown in Fig.~\ref{TDFTEO} in Appendix~\ref{AA}, indicating that it is characteristic to the GaP/Si interface.  The LFM cannot be a microscopic interface (MIF) mode arising from a ``wrong bond" \cite{Zhang1996, Jin1992}  such as Ga-Si or an interface mode arising from electrostatic effect at the interface \cite{Sood1985, Meynadier1987, Popovic1989a, Popovic1989b, Mowbray1991}, because they should appear at much higher ($>10$~THz) frequencies.

To quantitatively evaluate the LFM we fit the oscillatory signals in Fig.~\ref{Parameters}(a) to a damped harmonic function:
\begin{equation}\label{dh}
f(t)=\sum_i A_i \exp(-\Gamma_i t) \sin(2\pi\nu_i t+\phi_i),
\end{equation}
with $i$ denoting different phonon modes.  The obtained LFM amplitude $A_\text{LFM}$, dephasing rate $\Gamma_\text{LFM}$, frequency $\nu_\text{LFM}$, and initial phase $\phi_\text{LFM}$ are plotted as a function of $d$ in Fig.~\ref{Parameters}(a).  

The LFM frequency of all the samples falls within $\nu_\text{LFM}=2.0\pm0.2$~THz, though the frequency for sample A ($d=8$~nm) is somewhat lower than the others [Fig.~\ref{Parameters}(d)].   The dephasing rate [Fig.~\ref{Parameters}(e)] is not systematically dependent on $d$, except for the apparently smaller dephasing for sample F. 
The overall insensitivity of the frequency and dephasing rate to the growth stage confirms that the atomic bonding that gives rise to LFM is robust and its strength is hardly affected by the high-temperature overgrowth.  

The dependence of the LFM amplitude on the growth stage [filled symbols in Fig.~\ref{Parameters}(a)] appears less straightforward.  Starting from the 8-nm thick nucleation layer (sample A), high-temperature overgrowth first reduces $A_\text{LFM}$ to nearly half ($d=18$~nm, sample B).  
Further overgrowth turns $A_\text{LFM}$ to a monotonic increase.  At the end of the overgrowth ($d=48$~nm, sample E) $A_\text{LFM}$ is almost four times larger than that of the nucleation layer (sample A). On the other hand, additional low-temperature nucleation on top of sample A  ($d=8$~nm) increases the LFM amplitude by nearly three times ($d=10$~nm, sample F), disproportionately to the increase in the GaP thickness.  The non-monotonic $d$-dependence of the amplitude does not trace the nearly monotonic $d$-dependence of $\Delta R_\text{max}$ plotted in Fig.~\ref{CarrierRegA}(b).

\subsection{Optical polarization-dependence}\label{3B}

\begin{figure}
\includegraphics[width=0.475\textwidth]{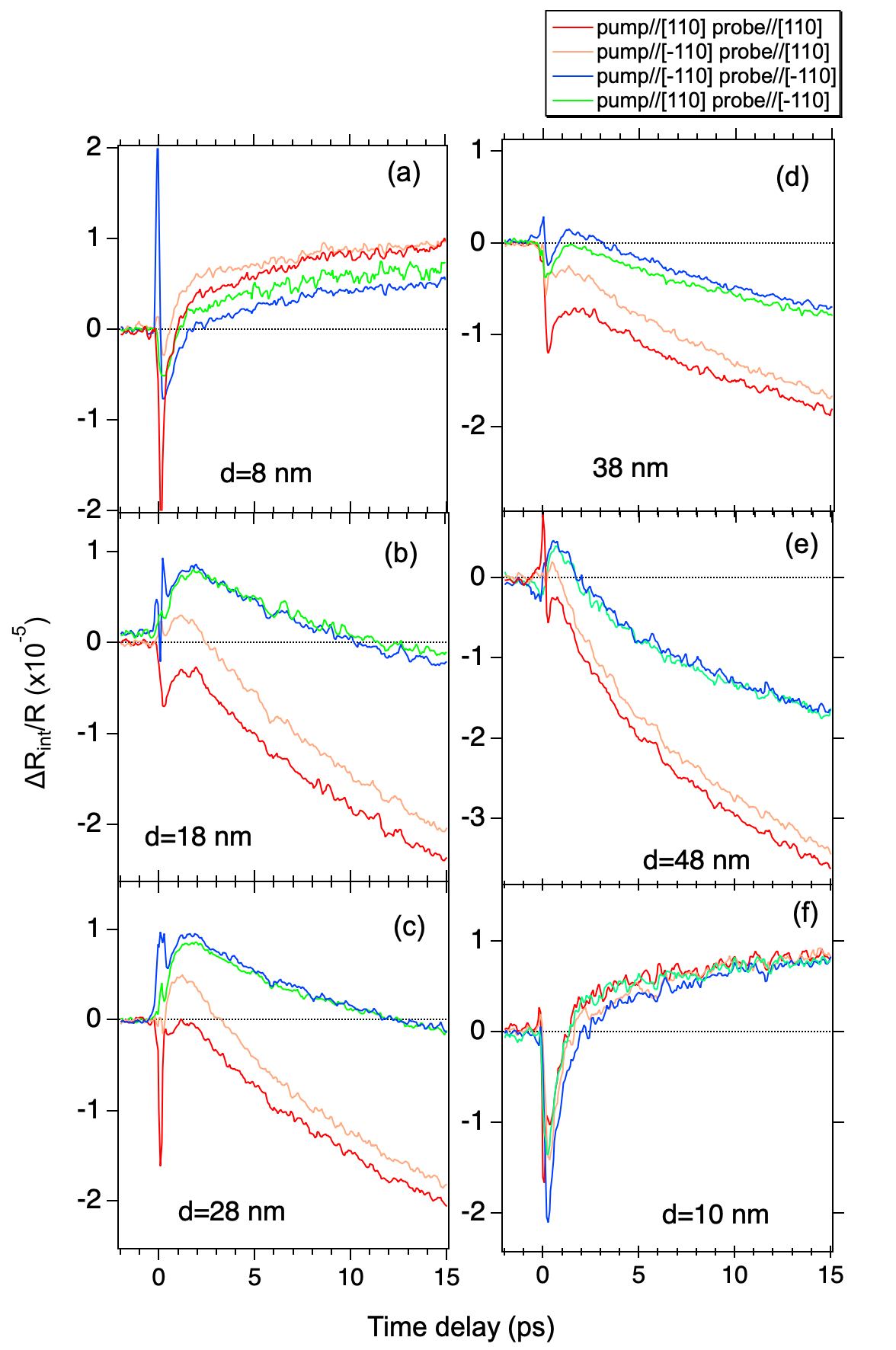}
\caption{\label{PolSlow} Interface contribution to the slow-scan TR signals obtained from GaP/Si samples A (a), B (b), C (c), D (d), E (e), and F (f) pumped and probed at 810 nm with four different pump/probe polarization combinations.}
\end{figure}

In Sect.~\ref{3A} the pump and probe polarizations were fixed to be parallel to the [110] or miscut direction of the Si substrate.  We now vary the pump/probe polarizations and see the effect on the carrier and phonon dynamics.  Rotating the pump and/or probe polarization by 90$^\circ$ from the [110] hardly affects the slow-scan TR signals of the nucleation layers (samples A and F), as shown in Fig.~\ref{PolSlow}(a) and (f).  For the overgrown layers  (samples B to E), by contrast, the initial drop height $\Delta R_\text{max}$ is consistently largest when the pump and probe are both polarized along the [110] direction of the substrate, as shown in Fig.~\ref{PolSlow}(b-e), though a coherent spike at $t=0$ appearing with the parallel polarization combination makes quantitative evaluation difficult.  Interestingly, the TR signals for the overgrown layers depend significantly on the \emph{probe} polarization but not so much on the pump.

\begin{figure}
\includegraphics[width=0.475\textwidth]{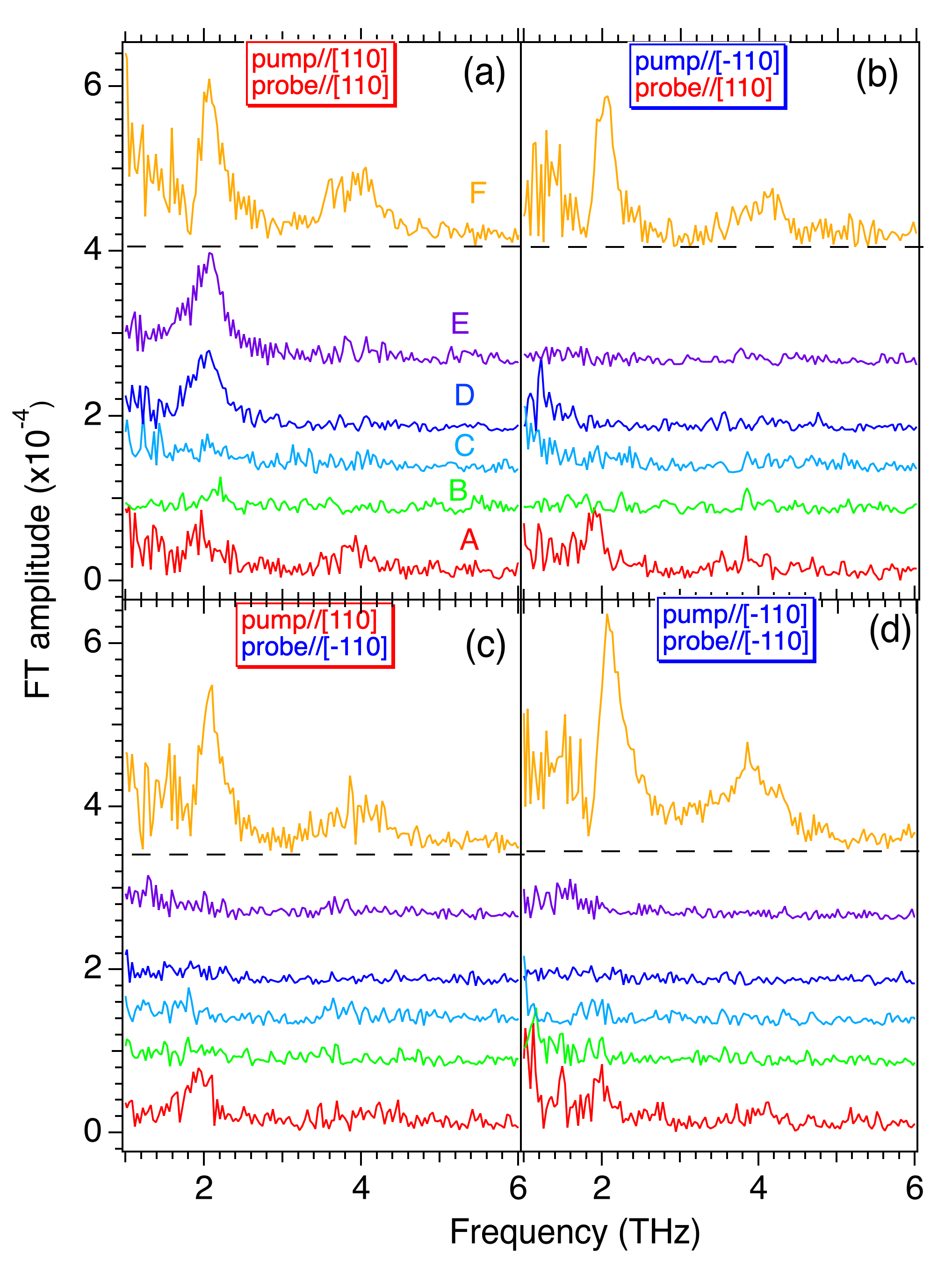}
\includegraphics[width=0.4\textwidth]{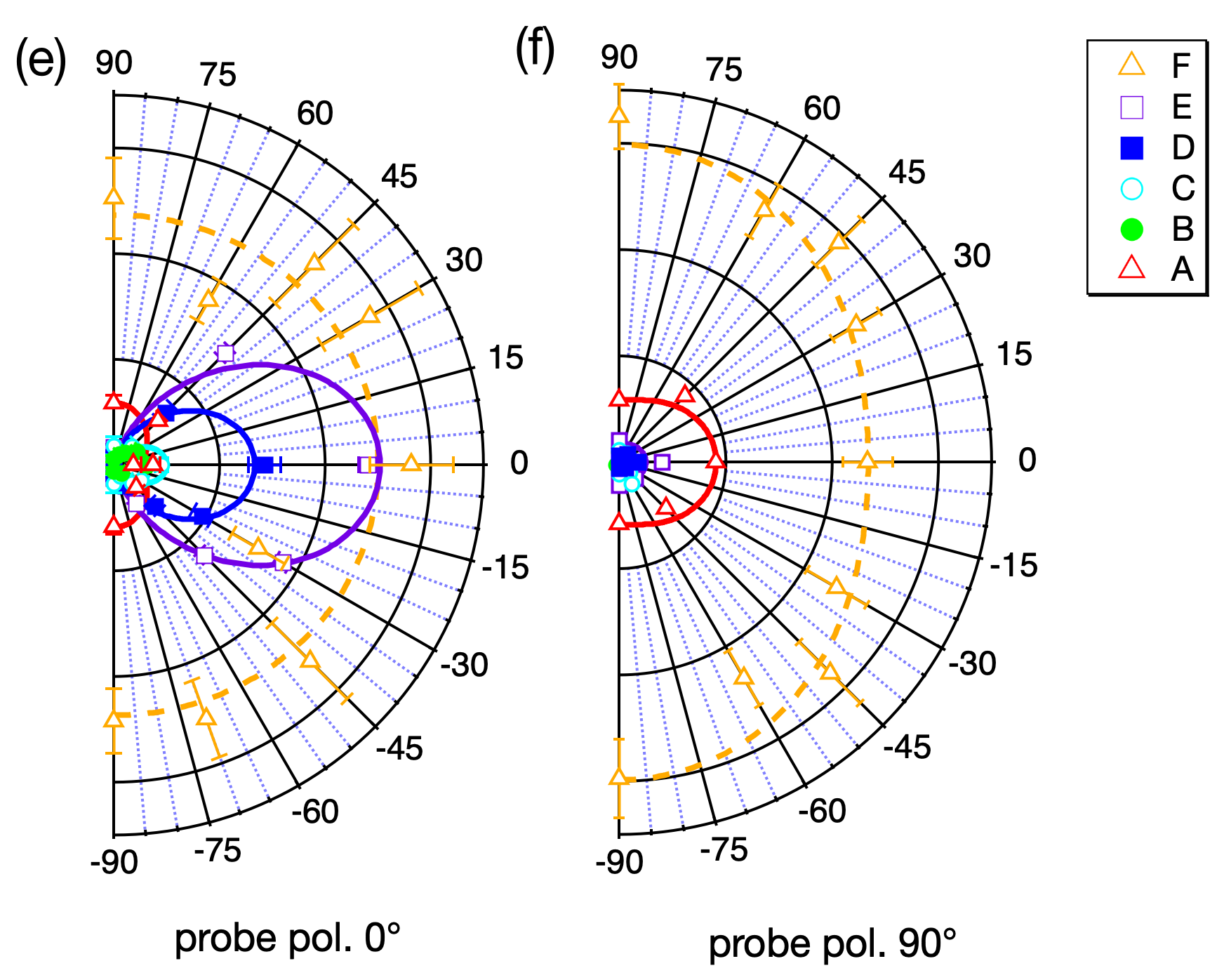}
\caption{\label{Pol} (a-d) FFT spectra of the oscillatory fast-scan TR signals from samples A to F pumped and probed at 815 nm with four different pump/probe polarization combinations.  Traces are offset for clarity. (e,f) Polar plot of the LFM amplitude as a function of pump polarization angle relative to the [110] direction of the Si substrate (symbols).   Probe polarization is along the [110] and [-110] directions in (e) and (f).  Curves represent fits to Eq.~(\ref{sinu}).
}
\end{figure}
  
Figure~\ref{Pol}(a-d) compares the FFT spectra of the oscillatory TR signals obtained at different pump-probe polarization combinations in the fast-scan scheme.  We find that the LFM peak amplitude of the overgrown layers (samples B to E) is reduced significantly by rotating the pump and/or probe polarization by 90$^\circ$ from the [110].  For the nucleation layers (A and F), by contrast, the peak amplitude is only moderately affected  by the same procedure.  The susceptibility of the overgrown layers to the pump/probe polarizations and the insensitivity of the nucleation layers are consistent, though not in quantitative agreement, with what we have seen for the carrier dynamics in Fig.~\ref{PolSlow}.
We further perform TR measurements at intermediate pump polarization angle and obtain the LFM amplitude, whose results are summarized in Fig.~\ref{Pol}(e,f) for two differently polarized probes.  In both panels $A_\text{LFM}$ can be fitted to:
\begin{equation}\label{sinu}
A_\text{LFM}(d,\theta)=C(d)+B(d) \cos2\theta,
\end{equation}
with $\theta$ being the angle of the pump polarization relative to the [110] direction.
With the $\theta$-dependence obtained from the fitting we can make a correction to the LFM amplitude plotted with the filled symbols in Fig.~\ref{Parameters}(c), whose procedure is descried in Appendix \ref{AC}.  The result, shown with open symbols in the same figure, emphasizes the amplitude of the nucleation layers in comparison with that of the overgrown layers but still exhibits a non-monotonic trend that is qualitatively similar to that before the correction.

\subsection{Pump wavelength-dependence}\label{3C}

\begin{figure}
\includegraphics[width=0.475\textwidth]{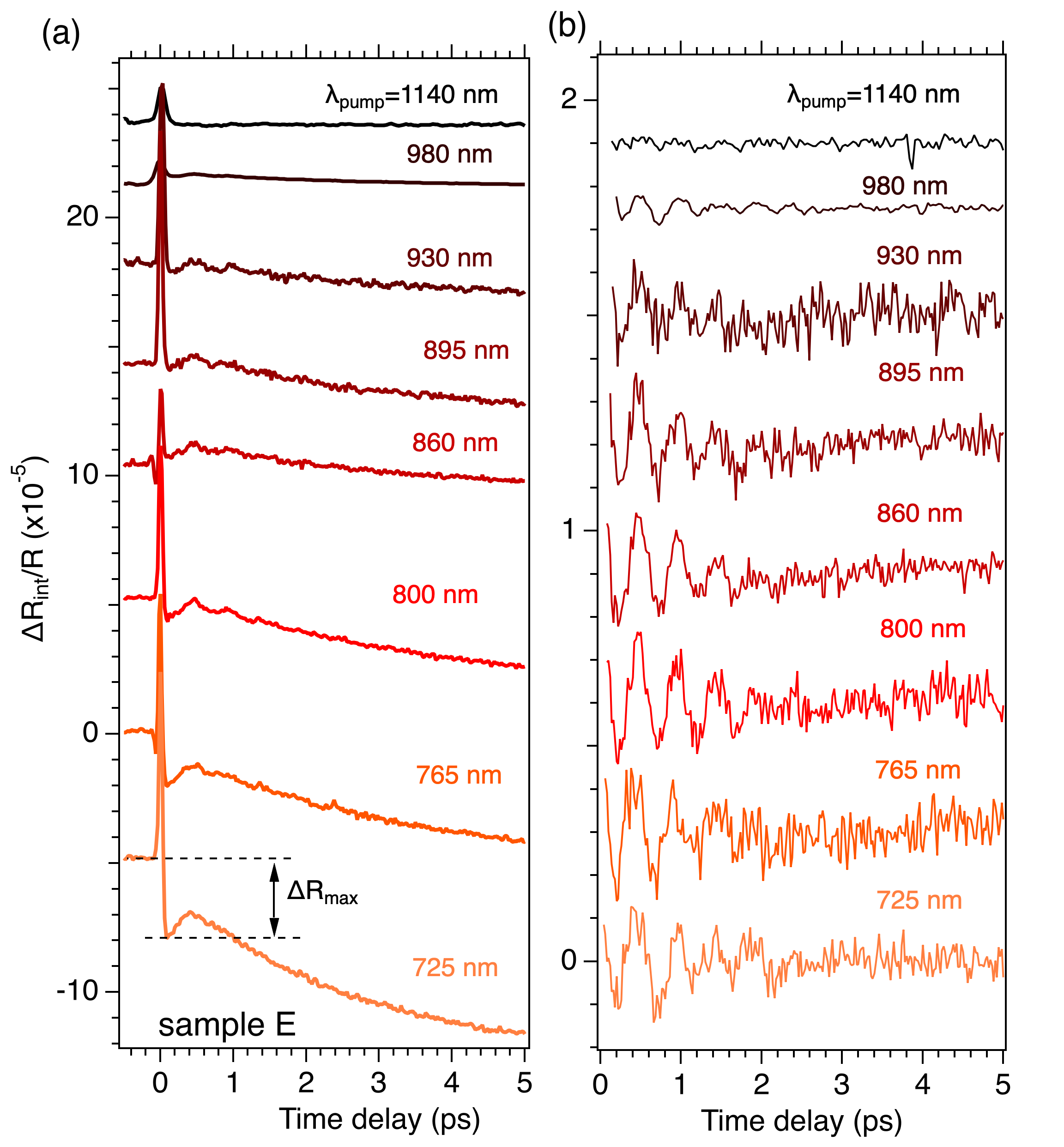}
\includegraphics[width=0.475\textwidth]{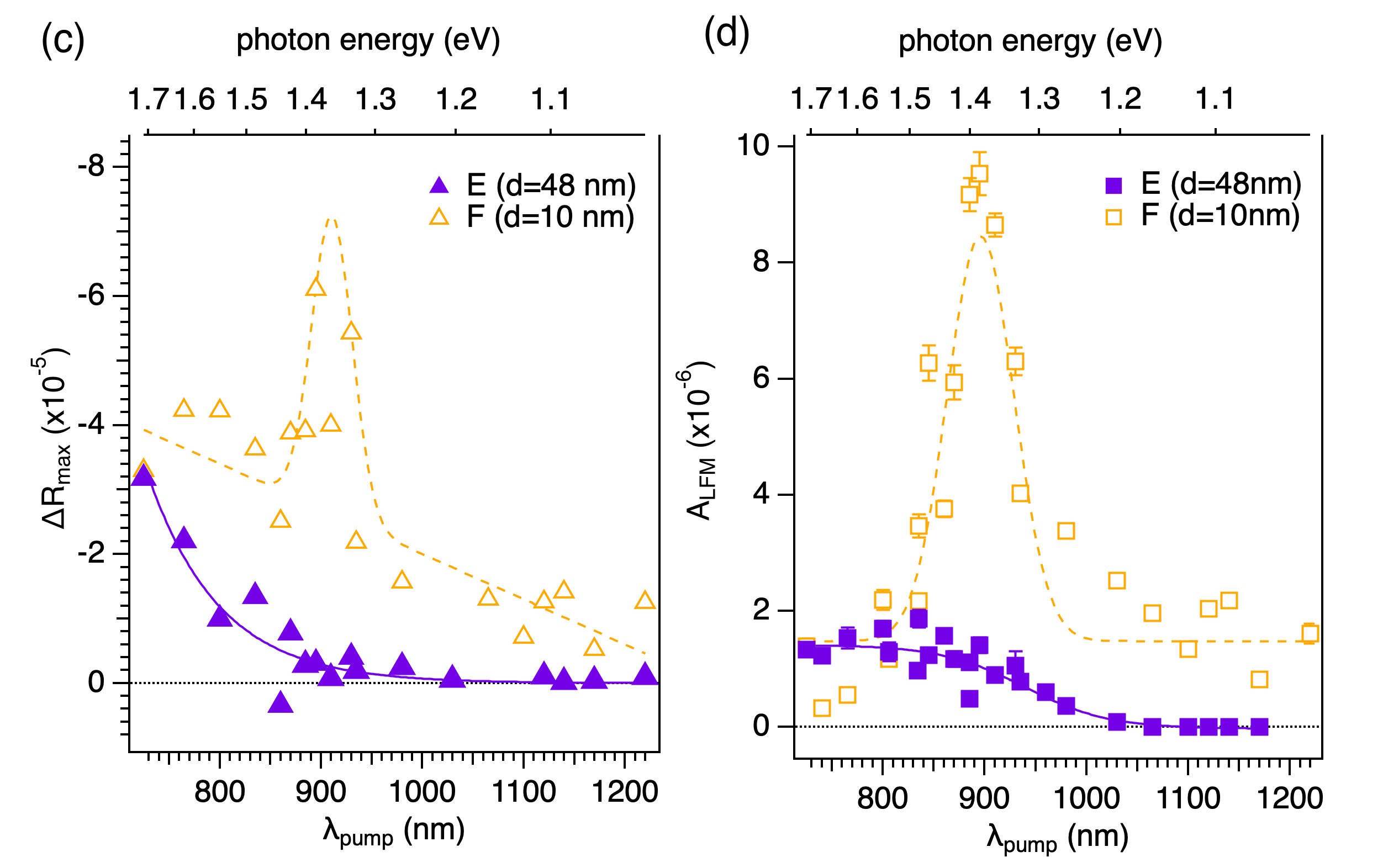}
\caption{\label{2color} (a,b) Interface-contribution to TR signal  (a) and its oscillatory component (b) obtained from sample E pumped at different wavelength and probed at 800 nm.  Incident pump fluence is $\sim0.4$~mJ/cm$^2$.  Traces are offset for clarity.  
(c) Initial drop height of the interface TR signal in (a)  as a function of pump wavelength (filled symbols).  (d) LFM amplitude obtained from fitting the oscillatory signal in (b) to Eq.~(\ref{dh}) (filled symbols).  Open symbols in (c) and (d) represent the corresponding quantities obtained from sample F.  Curves are to guide the eye.   
} 
\end{figure}

Sects.~\ref{3A} and \ref{3B} have revealed that the low-temperature nucleation layers (samples A and F) and the high-temperature overgrown layers (B to E) exhibit contrasted interface carrier and phonon dynamics, while within each growth categories the dynamics appears qualitatively similar.  In this subsection we compare the resonance behavior of the nucleation layer and the overgrown layer by performing two-color pump-probe experiments on one representative sample from each category.  

Figure~\ref{2color}(a) compares the interface contribution $\Delta R_\text{int}$ to the TR signal obtained from sample E pumped at different wavelengths and probed at 800 nm.  The corresponding TR signals from the nucleation layer sample F is shown in Fig.~1 of Ref.~\onlinecite{Mette2025}.  The TR signals from both samples exhibit an instantaneous drop at $t=0$, whose height $\Delta R_\text{max}$ is plotted against pump wavelength $\lambda_\text{pump}$ in Fig.~\ref{2color}(c).   $\Delta R_\text{max}$ from the nucleation layer shows a distinct resonance peak at $\lambda_\text{pump}=900$~nm (photon energy of 1.4~eV), which was attributed to an optical transition at the heterointerface involving a discrete electronic state \cite{Mette2020, Mette2025}, as illustrated with a vertical arrow in Fig.~\ref{banddiagram}(b).
Surprisingly, $\Delta R_\text{max}$ from the overgrown layer decreases monotonically with increasing $\lambda_\text{pump}$ without showing a distinct resonance peak, as plotted with filled symbols in Fig.~\ref{2color}(c).  
The absence of the resonance peak implies that the discrete interface electronic state of the nucleation layer is quenched by the high-temperature overgrowth.  The interface carrier dynamics of the overgrown layer is dominated most likely by charge transfer from the Si substrate into the GaP layer, as illustrated in Fig.~\ref{banddiagram}(c), as we have speculated in Sect.~\ref{3A1}. 

The TR signals from both samples also exhibit coherent LFM oscillation, as extracted in Fig.~\ref{2color}(b) for the overgrown layer and in Fig. 1 in Ref.~\onlinecite{Mette2025} for the nucleation layer. The LFM amplitude obtained by fitting  to Eq.~(\ref{dh}) is plotted as a function of $\lambda_\text{pump}$ in Fig.~\ref{2color}(d).   Whereas $A_\text{LFM}$ for the nucleation layer (sample F, $d=10$~nm) shows a distinct peak at $\lambda_\text{pump}=900$~nm, that for overgrown layer (sample E, $d=48$~nm) decreases monotonically with increasing pump wavelength.   The obtained resonance behavior traces that of $\Delta R_\text{max}$ of the respective sample in Fig.~\ref{2color}(c), which supports strong coupling between the LFM and the interface electronic transition regardless of the presence/absence of a discrete electronic state. 

\section{discussion}\label{disc}

The observation of the 1.4-eV resonance peak in $\Delta R_\text{max}$ for the nucleation layer and its absence for the overgrown layer [Fig.~\ref{2color}(c)] indicates that the discrete electronic state at the interface of the nucleation layer is quenched by the high-temperature overgrowth.  
The quench may be induced by reorganization in the interfacial atomic configuration. A previous DFT study \cite{Romanyuk2016} predicted that the atomically abrupt GaP/Si interface is less energetically favorable than intermixed interfaces, and that the energetic positions and densities of the interface electronic states are susceptible to the degree of the intermixing.
We therefore speculate that the low-temperature nucleation layer has a more abrupt interface that hosts the discrete electronic state.   High-temperature overgrowth reorganizes the interface to a more intermixed one, which leads to an apparent quench of the discrete interface state.

\begin{figure}
\includegraphics[width=0.475\textwidth]{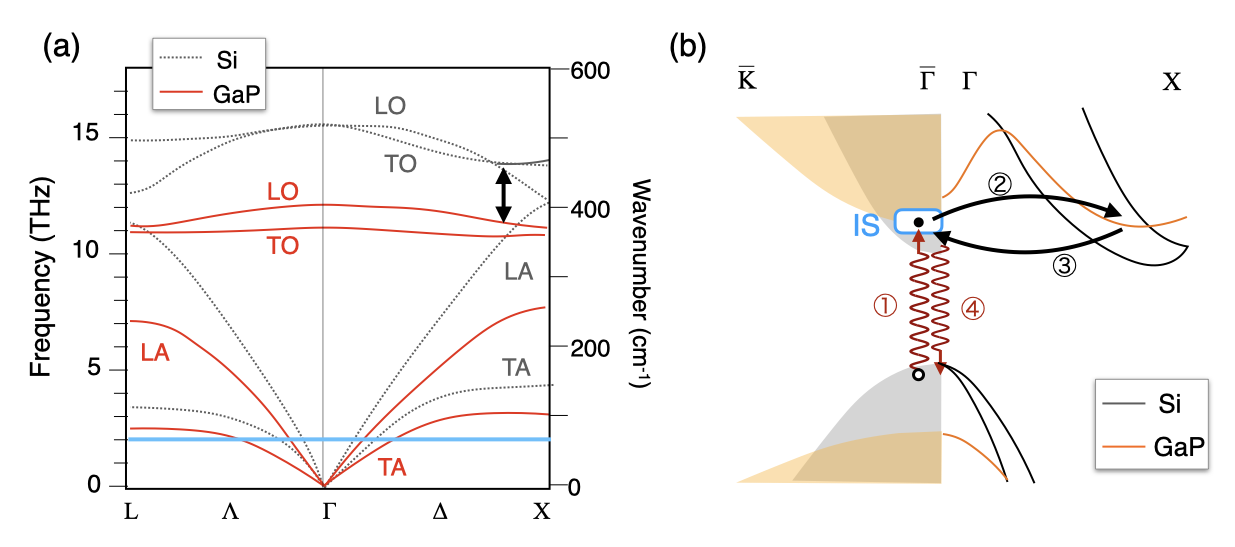}
\caption{\label{Diagram}
(a) Phonon dispersion relations of GaP (red solid curves) and Si (grey dotted curves) with an example of a pair of Si and GaP optical phonons that can give rise to a difference combination mode at 2~THz (black arrow).  Blue horizontal line denotes the frequency of 2~THz.  (b) Schematic illustration of a second-order Raman scattering involving large-wavevector Si-like and GaP-like phonons (steps 1 to 4 described in the main text) in the presence of an unoccupied interface electronic state (IS). 
Black and orange solid curves represent the bulk electronic bands of Si and GaP; grey and orange shades are their projection to the (001) plane.
}
\end{figure}

The LFM, by contrast, is detected at 2~THz for all the GaP/Si samples examined in the present study.  The constant frequency indicates the local atomic bond strength to be hardly affected by the interface reorganization during the high-temperature overgrowth.  Whereas the frequency of 2 THz falls within the acoustic continua at finite wavevectors ($q\neq0$) in the phonon dispersion curves of GaP and Si \cite{Eckl1996, Giannozzi1991} shown in Fig.~\ref{Diagram}(a), the LFM is unlikely to be a DATA or DALA mode of GaP, because they were observed as broad bands centered at 3 and 7~THz in previous Raman studies \cite{Malesh1989, Azhniuk2001}.  
In our previous TR study \cite{Mette2025} we proposed the LFM to be \emph{not} a fundamental phonon mode \emph{but} a difference combination mode between a GaP-like and a Si-like optical phonons, in analogy to the difference frequency Raman scattering from GaAs/AlAs superlattices \cite{Spitzer1992}.  
An example for such a phonon pair is indicated in Fig.~\ref{Diagram}(a), and the corresponding difference frequency scattering is illustrated schematically in Fig.~\ref{Diagram}(b).  The scattering consists of (1) excitation of an electron from a projected Si valence band to an unoccupied interface electronic state (IS), (2) scattering of the electron to the $X$ valley by absorbing a GaP-like optical phonon with a large wavevector $\bm{q}$, (3) scattering of the electron back to the projected conduction band by emitting a Si-like optical phonon with wavevector $-\bm{q}$, and (4) the electron-hole pair recombination yielding a photon whose energy is shifted by the difference between the Si-like and the GaP-like phonons.
When the incident photon energy is resonant to that of the transition to the IS in step 1, the entire process can be regarded as a triply resonant second-order Raman scattering \cite{Liu1991, Liu1992},  which could be orders-of-magnitude more efficient than non-resonant second-order scattering. 
Even if the incident photon energy is off the resonance or the IS is quenched, the triply resonant second-order scattering process can still take place with the projected GaP conduction band acting as an intermediate state in step 1, possibly with a different resonance behavior than in the presence of the IS.  

Coherent excitation of the difference combination mode described above would require a short ($<50$~fs) laser pulse because it involves coherent excitation of high-frequency ($\geq10$~THz) optical phonons.  Indeed, when we employed pump and probe pulses with longer duration in Figs.~\ref{CarrierRegA} and \ref{PolSlow} we detected no LFM.
Furthermore, in our previous TR experiments using NUV pulses with sufficiently short ($<10$~fs) duration \cite{Ishioka2017,Ishioka2019} we detected \emph{no} LFM from the GaP/Si interfaces but found the carrier and phonon dynamics to be dominated mostly by the above-bandgap excitation of bulk-like GaP and Si.   The absence of the LFM can be explained in terms of the intense direct transitions in GaP and Si overwhelming the much weaker indirect transitions involved in the second-order scattering process.

Whereas the model proposed above can account for the LFM frequency and the resonance behavior [Fig.~\ref{2color}(d)], the coupling with the interface carrier dynamics alone does not offer a quantitative explanation on the dependences on the pump/probe polarization [Fig.~\ref{Pol}] and on the growth stage [Fig.~\ref{Parameters}(c)].  Instead, the monotonic increase in the LFM amplitude by the overgrowth from $d=18$ to 48~nm may be suggesting the involvement of lattice imperfections within the GaP layer.  Anti-phase boundaries (APBs) \cite{Beyer2013, Belz2018, Beyer2019, Farin2019, Lenz2019}, which start from single-layer steps of the Si(001) substrate and mark boundaries between the main phase and the antiphase of GaP [Fig.~\ref{AFM}(g)], can be a candidate for such an imperfection.  The microscopic mechanisms of how the APBs can give rise to the LFM is unclear from the present results, however, as we surmise in Appendix~\ref{AD}.  Further STEM analyses on the GaP/Si heterointerfaces are currently under planning to investigate the possible role of the APBs and other lattice imperfections in the optical excitation of the LFM.

\section{conclusion}

We have systematically investigated the influence of high-temperature overgrowth on the ultrafast electron–phonon coupling dynamics at a GaP/Si(001) interface under below-bandgap excitation of GaP. Our results reveal that a discrete electronic state  observed in the low-temperature nucleation layer is effectively quenched following high-temperature overgrowth, likely due to atomic reorganization leading to a more intermixed interface.
Despite these substantial electronic modifications, a 2-THz phonon mode that is characteristic to the GaP/Si interface remains remarkably robust against high-temperature processing.  Its resonance behavior undergoes pronounced changes, reflecting its sensitivity to the altered interfacial electronic transitions with which it interacts. This highlights the intricate coupling between lattice vibrations and electronic structure at buried interfaces.
Our results show that lattice dynamics at buried interfaces can persist despite substantial electronic restructuring, while remaining highly sensitive to changes in interfacial states. More broadly, our work demonstrates that transient reflectivity provides direct access to coupled electronic and vibrational dynamics at buried heterointerfaces. This approach offers valuable insight for the design and optimization of heterostructures where interface quality plays a critical role in device performance.

\begin{acknowledgments}
The authors thank NIMS RCAMC and NAsP III/V GmbH for AFM measurements.  
We gratefully acknowledge funding by the Deutsche Forschungsgemeinschaft through the SFB 1083.  CJS was partially supported by the Air Force Office of Scientific Research under Award No. FA9550-24-1-0059.

\end{acknowledgments}

\appendix

\section{TR signals of bulk semiconductors}\label{AA}

\begin{figure}[H]
\includegraphics[width=0.475\textwidth]{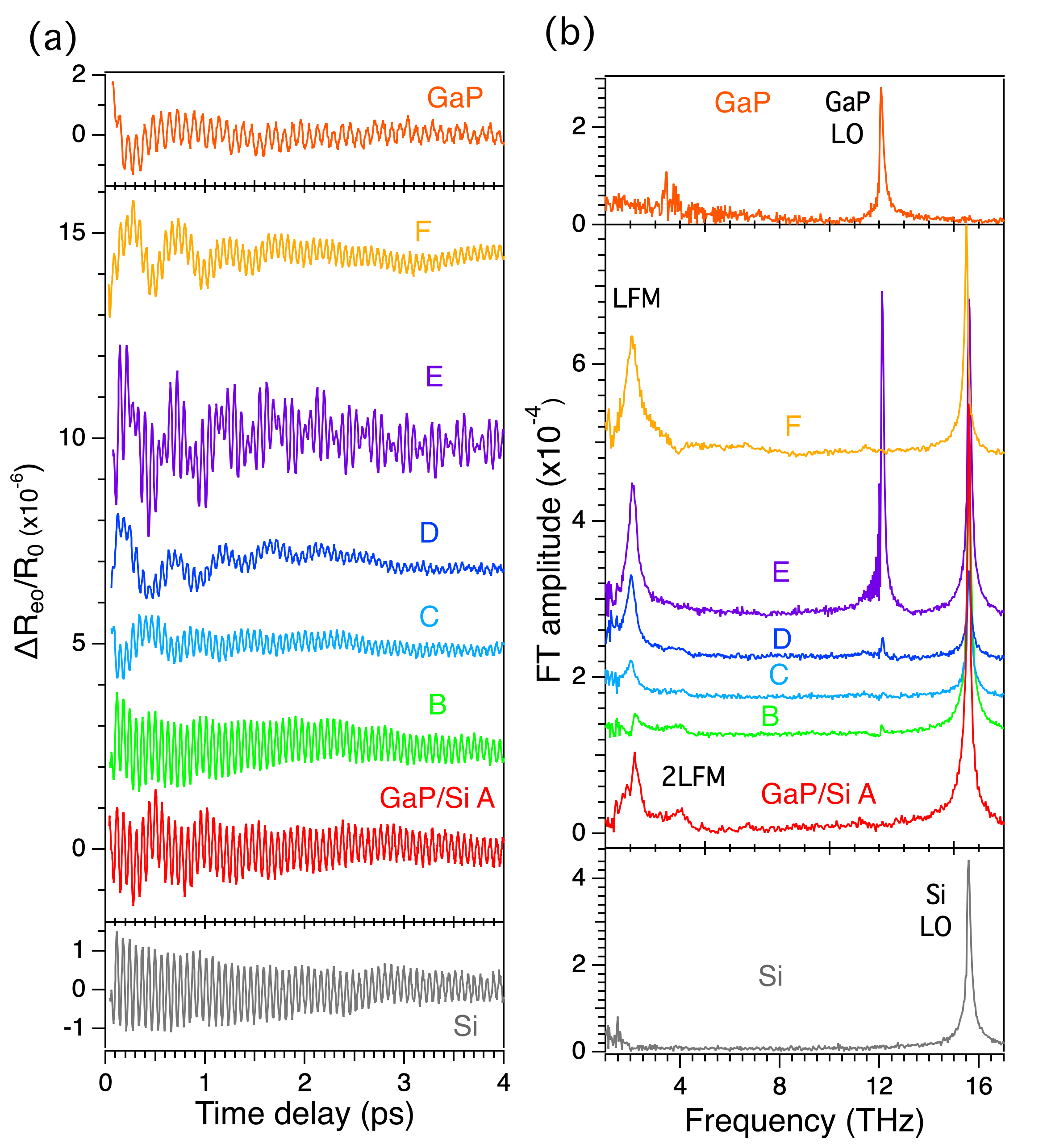}
\caption{\label{TDFTEO} Oscillatory part of anisotropic transient reflectivity (a) and its FFT spectrum (b) obtained from GaP/Si samples A to F and from (001)-oriented Si and GaP wafers.  Pump and probe wavelength is 815 nm.  Pump polarization is parallel to the [110] direction of the Si substrate, and probe light is polarized nearly along the [100] direction for the anisotropic detection.   Incident pump density is 0.18 mJ/cm$^2$.  Traces from GaP/Si samples are offset for clarity. The feature at $\sim$3.5 THz for bulk GaP in (b) is not reproducible and attributed to a noise.}
\end{figure}

Figure~\ref{TDFTEO}(a) compares the oscillatory part of the anisotropic reflectivity change $\Delta R_{eo}\equiv\Delta R_H-\Delta R_V$ obtained from blank Si and GaP  wafers with that of GaP/Si samples pumped and probed at 815 nm.  Here the anisotropic reflectivity change is measured by detecting the horizontal ($H$) and vertical ($V$) polarization components of the reflected probe light with a pair of matched photodiode detectors.  In contrast to the TR signals from the GaP/Si saimples, those from bulk GaP and Si exhibit no oscillatory feature at 2 THz, as is obviously seen in the FFT spectrum in Fig.~\ref{TDFTEO}(b).  
The absence of the 2-THz peak is in contrast to the GaP/Si samples but consistent with earlier Raman studies on the bulk semiconductors \cite{Hoff1973, Temple1973}. 

\section{Si substrate contribution to transient reflectivity of GaP/Si interface}\label{AB}

\begin{figure}
\includegraphics[width=0.475\textwidth]{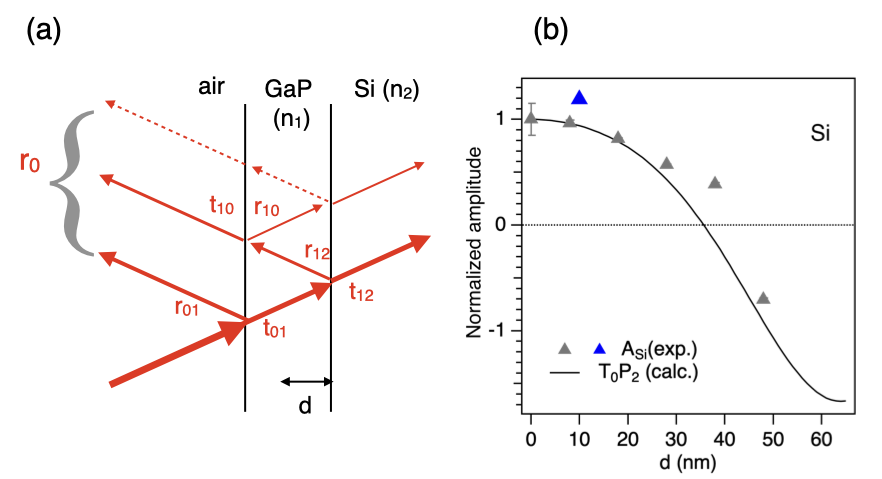}
\caption{\label{A_vs_d}  (a) Schematic illustration of the transmission and reflection of probe electric field that is incident on a GaP/Si interface.  $r_{ij}$ and $t_{ij}$ denote amplitude ratio of reflection and transmission for the light coming from $i$-th medium to the $j$-th medium ($i=0$: air, 1: GaP, and 2: Si).  Incidence angle is exaggerated for clarity.  (b) $d$-dependence of the coefficient $T(d)P_2(d)$ in Eq.~(\ref{sub}) (solid curve).  
Filled symbols represent the $d$-dependence of Si LO phonon amplitude obtained from the oscillatory TR signals in Fig.~\ref{TDFTEO}(a).  
}
\end{figure}

We consider the transmission and reflection of probe light at a GaP/Si interface consisting of a GaP layer of thickness $d$ on top of semi-infinitely thick Si substrate.  We assume a light pulse with frequency $\omega_0$ and wavevector $k_0=\omega_0/c$ is incident from the normal direction.  At the air/GaP interface, the light pulse is partially reflected and partially transmitted into the GaP layer with amplitude ratio $r_{01}$ and $t_{01}$, as schematically illustrated in Fig.~\ref{A_vs_d}(a).  Likewise, the light wave transmitted into the GaP layer is partially reflected at the GaP/Si interface and partially transmitted into the Si substrate with amplitude ratio $r_{12}$ and $t_{12}$.  After the reflection at the GaP/Si interface, the light wave is again partially reflected at the GaP/air interface and partially transmitted into air, with amplitude ratio $r_{10}$ and $t_{10}$.  While the light wave crosses the GaP layer once it gains a thickness-dependent phase shift:
\begin{equation}
\phi\equiv{n}_1k_0d,
\end{equation}
with $n_1$ denoting the refractive index of GaP.  The amplitude ratio $r_0$ of the outgoing wave into air to the incident wave can be given by the sum of multiple reflection pathways: 
\begin{eqnarray}\label{th8}
r_0=\dfrac{r_{01}+r_{12}e^{2i\phi}}{1+r_{01}r_{12}e^{2i\phi}}.
\end{eqnarray}
Here
\begin{eqnarray}\label{th3}
r_{01}=\dfrac{1-{n}_1}{1+{n}_1}; \;\; r_{12}=\dfrac{{n}_1-{n}_2}{{n}_1+{n}_2}.
\end{eqnarray}
with $n_2$ denoting the refractive index of Si.  
The reflectance, or the light intensity reflected into air, can be given by \cite{Ishioka2022}:  
\begin{eqnarray}\label{th10}
R_0=|r_0|^2=\dfrac{r_{01}^2+r_{12}^2+2r_{01}r_{12}\cos2\phi}{1+r_{01}^2r_{12}^2+2r_{01}r_{12}\cos{2\phi}}.
\end{eqnarray}
The transmittance, or the light intensity transmitted into the Si substrate, can be given by:  
\begin{eqnarray}\label{th11}
T_0=1-R_0=\dfrac{(1-r_{01}^2)(1-r_{12}^2)}{1+r_{01}^2r_{12}^2+2r_{01}r_{12}\cos{2\phi}}.
\end{eqnarray}

We assume that a separate pump wave induces small disturbances $\delta{n}_1$ and $\delta{n}_2$ in the refractive indices in the GaP layer and in the Si substrate and thereby modifies the reflectance:
\begin{equation}\label{th12}
\dfrac{\Delta R}{R_0}=\dfrac{1}{R_0}\left(\dfrac{\partial R_0}{\partial {n}_1}\delta {n}_1+\dfrac{\partial R_0}{\partial {n}_2}\delta {n}_2\right).
\end{equation}
The  second term of Eq.~(\ref{th12}), representing the contribution of the Si substrate to the pump-induced reflectivity change, can be expressed by:
\begin{widetext}
\begin{eqnarray}\label{th15}
\dfrac{1}{R_0}\dfrac{\partial R_0}{\partial n_2}\delta n_2&=&\dfrac{2r_{12}(1-r_{01}^4)+2r_{01}(1+r_{12}^2)(1-r_{01}^2)\cos2\phi}{(r_{01}^2+r_{12}^2+2r_{01}r_{12}\cos2\phi)(1+r_{01}^2r_{12}^2+2r_{01}r_{12}\cos2\phi)}\times\dfrac{-2n_1}{(n_1+n_2)^2}\delta n_2\equiv P_2(d)\delta n_2.
\end{eqnarray}
\end{widetext}
Since the intensity of the transmitted light into Si, $T_0$,  also depends on the GaP layer thickness, the overall $d$-dependence of the Si substrate contribution is determined by that of the product $T_0 P_2$, which is shown with a solid curve in Fig.~\ref{A_vs_d}(b).   The calculation reasonably reproduce the experimentally obtained Si LO amplitude, which is plotted with filled symbols in the same figure.

Conversely, the $d$-dependence of the Si LO phonon amplitude can be used as a quantitative measure for the GaP overlayer thickness.  For example, the colored symbol in Fig.~\ref{A_vs_d} represents the Si LO amplitude from sample F ($d=10$~nm).  The experimentally obtained LO amplitude is in reasonable agreement with the theoretical prediction, confirming the thickness of sample F to be 10 nm despite the disproportionately intense LFM amplitude.

\section{Correction of polarization-dependence of LFM amplitude}\label{AC}

The LFM amplitude plotted in Fig.~\ref{Parameters}(c) was measured at a fixed pump polarization.  Fig.~\ref{Pol} revealed the pump polarization-dependence of the LFM was not identical for all the samples, however.  We therefore correct the LFM amplitude by taking into account its dependence on pump polarization angle $\theta$.  Open symbols in Fig.~\ref{Parameters}(c) represent $A_\text{LFM}$ integrated over $\theta$:  
\begin{equation}
\int _{-\pi/2}^{\pi/2}d\theta A_\text{LFM}(d,\theta)\simeq\int_{-\pi/2}^{\pi/2}d\theta C(d)+B(d) \cos2\theta.
\end{equation}
We find that both the increase in $A_\text{LFM}$ from $d=8$ to 10 nm and the decrease from $d=8$ to 18 nm are emphasized considerably after the correction.  As a result, $A_\text{LFM}$ for $d=10$~nm becomes twice as intense as that for $d=48$~nm.  Qualitatively, however, the $d$-dependence of the $A_\text{LFM}$ shows a similar, non-monotonic trend whether or not its pump polarization-dependence is taken into account.

\section{Possible explanation of LFM involving APBs}\label{AD}

In the phonon dispersion curve of GaP [red curves  in Fig.~\ref{Diagram}(a)] the frequency of 2~THz [blue line in the same figure] falls on the transverse acoustic (TA) phonon at about 1/3 the $\Gamma - X$ distance.  Such a TA phonon can also participate in an electron intervalley scattering.
When scattering occurs between equivalent valleys along the same crystallographic axis, the process is referred to as a $g$-process, whereas scattering to an adjacent valley on a different axis is an $f$-process. These intervalley mechanisms can become particularly significant at phonon “hotspots” in semiconductor nanostructures \cite{Sinha2004, Ferry1976}.  A $g$-process requires a phonon wavevector $q\simeq0.3 (2\pi/a)$ in the $\langle 100 \rangle$ direction, i.e., about one-third of the $\Gamma - X$ distance.
Although the TA phonon is Raman inactive in a perfect GaP crystal, lattice imperfections such as antiphase boundaries (APBs) can introduce large wavevectors, rendering the $q\neq0$ phonons Raman active. Since the $g$ process involves phonons with a narrowly defined wavevector near the intervalley separation, this mechanism is also consistent with the relatively narrow linewidth of the low-frequency mode (LFM).

\bibliography{GaPSi_ref}

\end{document}